# Toward an artificial Mott insulator:

# Correlations in confined, high-density electron liquids in SrTiO$_3$

Pouya Moetakef[1], Clayton A. Jackson[1], Jinwoo Hwang[1], Leon Balents[2], S. James Allen[3], and Susanne Stemmer[1]

[1]Materials Department, University of California, Santa Barbara, California 93106-5050, USA

[2]Kavli Institute of Theoretical Physics, Santa Barbara, CA 93106-4030, USA

[3]Department of Physics, University of California, Santa Barbara, California 93106-9530, USA

**Abstract**

We investigate correlation physics in high-density, two-dimensional electron liquids that reside in narrow SrTiO$_3$ quantum wells. The quantum wells are remotely doped via an interfacial polar discontinuity and the three-dimensional (3D) carrier density is modulated by changing the width of the quantum well. It is shown that even at 3D densities well below one electron per site, short-range Coulomb interactions become apparent in transport, and an insulating state emerges at a critical density. We also discuss the role of disorder in the insulating state.

Strongly correlated ("Mott") insulators defy simple models that predict metallic transport. An early triumph of theory was to recognize that Coulomb repulsion can localize the conduction electrons, causing a magnetic insulator to emerge [1]. *Long-range* Coulomb interactions are important at low carrier densities typical for many semiconductors. *Short-range* interactions dominate if the probability for two electrons to occupy a single site becomes significant, which requires high electron densities [2-4], and result in metal-to-insulator transitions in materials with narrow bands. The pathways between this insulating state and the conducting Fermi liquid, and associated phenomena, such as high-$T_c$ superconductivity, are subjects of intense study in condensed matter physics. In real materials, lattice distortions, charge and magnetic ordering often accompany the transition, making it difficult to isolate effects from electron correlations.

Oxide interfaces that exhibit a polar discontinuity allow for two-dimensional electron liquids (2DELs) with unprecedented charge densities [5, 6]. For example, the $GdTiO_3/SrTiO_3$ interfaces shown in Fig. 1 produce a 2DEL with ½ electron per interface unit cell, a sheet density of ~$3.3\times10^{14}$ cm$^{-2}$ per interface [7]. Here, such extreme electron densities are introduced into the *d*-bands of $SrTiO_3$ (see ref. [7]), a material that is a band insulator in bulk. In heterostructures that contain two such interfaces, carrier concentrations in sufficiently narrow $SrTiO_3$ quantum wells may approach one electron per site and the conditions for Hubbard-type strong correlation physics. Because the interface dopes the quantum well [8], high electron densities are obtained without introducing dopant atoms, which would cause disorder and lattice distortions - a situation much closer to the ideal Hubbard-like models of many-body theory. By changing the width of the quantum well, the 3D electron density can be varied, which allows for a systematic study of interaction effects, without changing other parameters, and for addressing a number of scientific questions: (i) Do confined, high-density electron gases in $SrTiO_3$ exhibit signatures of short-range Coulomb interactions? For example, we recently reported weak ferromagnetism (magnetoresistance hysteresis) in thin, high-density SrTiO$_3$ quantum wells [9], indicating correlation effects; however, because GdTiO$_3$ is ferrimagnetic, proximity effects could not be ruled out as the source of the magnetism [9]. (ii) Can a low-dimensional, artificial Mott insulator (AMI) with a correlation-induced gap be created by purely electrostatic doping of a band insulator? This Rapid Communication addresses both questions and also discusses the role of disorder in the insulating state.

All samples were grown by molecular beam epitaxy, as described elsewhere [7, 10]. Structures included GdTiO$_3$/SrTiO$_3$ grown on (001) (LaAlO$_3$)$_{0.3}$(Sr$_2$AlTaO$_6$)$_{0.7}$ (LSAT) substrates, containing either a single electrically active interface (GdTiO$_3$/SrTiO$_3$/LSAT) or two active interfaces (GdTiO$_3$/SrTiO$_3$/GdTiO$_3$/LSAT), as well as Gd$_x$Sr$_{1-x}$TiO$_3$ films grown on LSAT or SrTiO$_3$ single crystals, respectively. Details of the growth conditions and methods, as well as structural, electrical, and magnetic characterization can be found elsewhere [7, 9, 10]. Ohmic contacts for LSAT/GdTiO$_3$/SrTiO$_3$/GdTiO$_3$ and LSAT/SrTiO$_3$/GdTiO$_3$ were 50 nm-Ti/300-nm-Au, while for Gd$_x$Sr$_{1-x}$TiO$_3$ structures the contacts were 40-nm-Al/20-nm-Ni/300 nm Au (with Au being the top layer). Temperature dependent electrical and magnetotransport properties were measured using Van der Pauw geometry in a physical properties measurement system (Quantum Design PPMS), as described in detail elsewhere [7, 9, 10].

The thickness of the SrTiO$_3$ quantum wells was varied from a few-nm to one unit cell (u.c.), which changes the 3D density. For Mott transitions, the electron count per Ti site is important, thus the quantum well thickness notation used here should be clarified. Growth of the GdTiO$_3$ layers was started/ended with a TiO$_2$ layer. The SrTiO$_3$ thicknesses given here are those derived from the growth rate of the SrTiO$_3$ layers, and do not include the interfacial TiO$_2$ layers. The reported 3D carrier concentrations (*n*) are calculated by dividing the sheet carrier density by that thickness. For example, a quantum well of one u.c. (or 0.4 nm thickness), shown in Fig. 1(a), consisted of two SrO layers bound by two $TiO_2$ layers, each of which is shared with the adjacent $GdTiO_3$. Quantum wells labeled 0.8 nm consisted (on average) of 3 SrO layers (2 u.c.), bound by two $TiO_2$ layers shared with the $GdTiO_3$, and so forth. Furthermore, for quantum wells thicker than 4 nm it was assumed that most of the electrons in the 2DEL were confined to the first 4 nm. Support for this assumption comes from the fact that the mobility increased only moderately for wells thicker than 4 nm, and, as a result, the sheet resistance barely changed [11], and from the coefficient for the quadratic temperature dependence of the resistivity, discussed below. Theoretical calculations indicate that for 2DELs in $SrTiO_3$, containing $3.5 \times 10^{14}$ cm$^{-2}$ carriers, about 75% of the carriers are confined within the first three layers [12]. Returning to the question of electron count per Ti site, the thinnest quantum well studied here, a single u.c. of $SrTiO_3$ embedded in $GdTiO_3$, accommodate ~ $7 \times 10^{14}$ cm$^{-2}$ carriers [11], or about one electron shared between three $TiO_2$ layers, if the two interfacial $TiO_2$ layers are included in the electron count. However, the staggered band alignment [7] favors charge transfer into the $SrTiO_3$ well, and the precise electron distribution over the three $TiO_2$ layers likely will depend on a number of atomistic parameters, including atomic relaxations [13, 14].

High-angle annular dark-field scanning transmission electron microscopy (HAADF-STEM) images [Fig. 1(b)] confirmed that continuous, $SrTiO_3$ quantum wells down to one u.c. in thickness were obtained. The excellent control over quantum well thicknesses afforded by MBE is also apparent in the sheet resistance data of these samples [11]. For example, changing the thickness by one monolayer produced detectable decreases in the sheet resistance [11], due to the increase in mobility (interface roughness scattering in thin quantum wells scales with ~ $a^6$, where *a* is the quantum well thickness [15]). Hall and Seebeck measurements indicated that transport occurred in the SrTiO$_3$, consistent with the band alignments in this system [7]. In particular, even for the thinnest quantum wells, the Hall and Seebeck coefficients were *n*-type, characteristic for SrTiO$_3$. In the Mott insulator GdTiO$_3$ these coefficients are *p*-type and positive [16]. Furthermore, mobilities were higher than those of doped GdTiO$_3$ for all samples.

All quantum well structures with SrTiO$_3$ thicknesses greater or equal to 2 u.c. ("0.8 nm") were metallic (Fig. 2). The upturn in resistivity at very low temperatures coincides with the appearance of negative magneto-resistance [11], indicative of weak localization [17]. As shown in Fig. 2(a), at temperatures above the onset of weak localization the temperature dependence of the resistivity, $\rho(T)$, can be described by Fermi-liquid theory as:

$$\rho(T) = \rho_0 + AT^2, \tag{1}$$

where $\rho_0$ is the residual resistivity due to impurity scattering, and *T* the temperature. The quadratic temperature dependence indicates electron-electron scattering as the dominant scattering mechanism [18] and was found for all samples investigated here, up to room temperature [11], except for the Gd$_{0.285}$Sr$_{0.715}$TiO$_3$ and Gd$_{0.56}$Sr$_{0.44}$TiO$_3$ films, which showed the quadratic temperature dependence only below 250 K.

The *A* coefficient in Eq. (1) is proportional to the mass enhancement due to electron correlations [18, 19], and other materials parameters, including the carrier concentration and band structure [20-22]. The quadratic dependence of *A* on the mass holds in both 2D and 3D [23]. In general, *A* decreases with increasing carrier concentration [18]. Figure 2(b) shows *A* as a function of *n* for quantum wells, Gd$_x$Sr$_{1-x}$TiO$_3$ thin films, and La$_x$Sr$_{1-x}$TiO$_3$ bulk samples from the literature [24, 25]. For all samples, $A$ decreases with increasing $n$ for $n \lesssim 7\times10^{21}$ cm$^{-3}$, consistent with Fermi liquid behavior. $A$ was constant for the quantum wells thicker than 4 nm (GdTiO$_3$/SrTiO$_3$/LSAT samples), consistent with the tight confinement of the 2DEL, discussed above. However, for the thinnest, still metallic quantum wells with the highest densities (0.8 and 1 nm thickness sandwiched between two GdTiO$_3$ layers), the behavior changes, and $A$ *increases* with $n$. This indicates that electron correlations are enhanced in the extreme 3D carrier concentration regime. This mass enhancement is also found in bulk La$_x$Sr$_{1-x}$TiO$_3$ (ref. [25]) and Gd$_x$Sr$_{1-x}$TiO$_3$ films (this study) on the verge to the transition to the Mott insulating state, i.e., at large filling factors $x$. A mass enhancement near the Mott metal insulator transition, as a half-filled band is approached, is predicted from Mott-Hubbard theory [4]. Theory also predicts $A$ to diverge near half-filling [26]. The similarity in the behavior of $A$ as a function of filling between bulk rare earth titanates near the filling-induced Mott metal-insulator transition, and the high carrier density quantum wells of SrTiO$_3$ is remarkable.

To the best of our knowledge the results are the first experimental demonstration that mass enhancement due to short-range, on-site Coulomb interactions (Mott-Hubbard physics) can be obtained in a two-dimensional band insulator, due to purely electronic effects, i.e., in the absence of chemical doping (alloying). Furthermore, the discontinuity in $A$ appears for all three materials (SrTiO$_3$ quantum wells, La$_x$Sr$_{1-x}$TiO$_3$, and Gd$_x$Sr$_{1-x}$TiO$_3$), at approximately the same 3D carrier density, $n_c \approx 7\times10^{21}$ cm$^{-3}$. This indicates similarity in the basic correlation physics in all three samples, and lends credibility to the way in which the 3D carrier concentration of the quantum wells was calculated from the thickness and sheet density (see above). The results suggest that increase in the $A$ coefficient occurs when the probability of double occupancy (and thus Hubbard-type strong correlation physics) becomes non-negligible, and that this depends *only* on density and not on the specific material (within this series). Thus, the effects of electrostatic or chemical doping are remarkably similar. Above this critical density, the *rate* of increase in *A* with filling appears to depend on the specific material. The electron-correlation strength depends on the rare earth titanate, specifically their bandwidth [27]. The correlation strength is larger in GdTiO$_3$ than in LaTiO$_3$; the results here show that this is correlated with $\partial A/\partial n$. By this measure, the correlation strength in the confined quantum wells appears to be intermediate between La$_x$Sr$_{1-x}$TiO$_3$ and Gd$_x$Sr$_{1-x}$TiO$_3$.

Even thinner quantum wells (0.4 nm) were insulating, although Hall measurements showed no significant change in carrier density [11]. Figure 3 shows the sheet resistance as a function of temperature for the insulating quantum wells with one u.c. SrTiO$_3$. The transition to an insulator for the high-density quantum well sandwiched between two GdTiO$_3$ layers appears to be consistent with a filling-controlled Mott metal-insulator transition. However, both disorder (Anderson) and on-site Coulomb interactions (Mott-Hubbard) can cause a transition to an insulating state. For a two-dimensional system with disorder and interacting electrons, theory predicts a universal cross-over between the correlated metallic state and an Anderson insulator near the Mott minimum conductivity, $\sim (e^2/h)$ or $\sim 25$ k$\Omega^{-1}$ (ref. [28]). The conductance of both samples was below this value. For extremely narrow quantum wells, and high electron densities, surface roughness scattering becomes the dominant mechanism of electron localization, relative to those due to impurities and remote dopant scattering [15]. Using the formalism developed in ref. [15], the interfacial roughness parameters for strong localization can be estimated for a given carrier density and quantum well thickness. A SrTiO$_3$ quantum well {in-plane mass of $\sim$ 0.9 $m_e$, out-of-plane mass $\sim$ 8 $m_e$, where $m_e$ is the free electron mass [29], a dielectric constant ($\varepsilon$) of 25 [30]}, with a thickness of 0.4 nm, and roughness parameters of $\Lambda = 15$ nm, $\Delta = 0.2$ nm, where $\Delta$ is the height and the $\Lambda$ the length of the Gaussian-like interface roughness, and a carrier density of $7\times10^{14}$ cm$^{-2}$, *is still metallic* [11]. The particular choice for $\Delta$ is justified by the HAADF images, but $\Lambda$ is more difficult to estimate. It is likely that both Anderson and Mott localization are at play here. In this case, the band tails are localized over a range of energy. Localized transport would occur by means of variable range hopping (VRH) in the band tail and thermal activation to a mobility edge. The resistivity in the VRH regime is described $\rho(T) \sim \exp(T_0/T)^y$, where $y$ is a parameter that has a value of ½ in case of strong electron interaction and the formation of a Coulomb gap in both 2D and 3D (Efros-Shklovskii law [31]). VRH gives way to Arrhenius behavior [$\rho(T) \sim \exp(\Delta E/kT)$, where $\Delta E$ is the activation energy and $k$ the Boltzmann constant] at higher temperatures, when carriers above the mobility edge participate in transport.

To determine the respective roles of disorder vs. electron correlations in the localization, data were fit to Arrhenius and VRH expressions, respectively. The GdTiO$_3$/1 u.c. SrTiO$_3$/GdTiO$_3$/LSAT sample, with a sheet carrier density of ~ $8\times10^{14}$ cm$^{-2}$, exhibits two distinct regimes at high and low temperatures, respectively. In particular, there is a clear crossover between Arrhenius ($\Delta E = 0.02$ eV, red dotted line in Fig. 3) and VRH with a Coulomb gap, $\rho(T) \sim \exp(T_0/T)^{1/2}$ (dashed black line in Fig. 3) with a characteristic temperature, $T_0$, of 8690 K. These results provide evidence of thermally activated transport to a mobility edge and band transport at high temperatures. The results support the interpretation of correlation modified density of states in which incoherent transport occurs, at the Fermi energy. At the same time, and similar to what has been reported for bulk titanates [27], the small activation energy appears to be a signature of close proximity to the metal-insulator transition.

For comparison, the insulating GdTiO$_3$/1 u.c. SrTiO$_3$/LSAT sample could be described over a large temperature range by the Efros-Shklovskii law (see Fig. 3), with a $T_0$ of 13750 K, indicating the importance of disorder. This quantum well suffers from more scattering due to the proximity to the substrate surface. Furthermore, as it contains half the 3D carrier density, the Fermi energy will be deeper in the band tail suppressing activation to the mobility edge.

The very large value for $T_0$ for both quantum wells gives further evidence for Hubbard physics. Efros-Shklovskii [31] relate $T_0$ to the localization length, $\xi$, assuming *only long-range Coulomb interactions*, $\xi = 6.5\left[e^2/(4\pi\varepsilon_0\varepsilon k T_0)\right]$, where $e$ is the electron charge, $\varepsilon_0$ is the permittivity of free space and the factor of 6.5 is for a 2D system [32]. Using $\varepsilon \approx 25$, we obtain $\xi = 0.5$ nm and $\xi = 0.31$ nm for the GdTiO$_3$/1 u.c. SrTiO$_3$/GdTiO$_3$/LSAT and GdTiO$_3$/1 u.c. SrTiO$_3$/LSAT, respectively, comparable to or smaller than a single unit cell. This clearly points to the importance of strong on-site Coulomb repulsion.

In summary, we have shown evidence for on-site Mott-Hubbard-type correlation physics in thin, high-carrier-density quantum wells of SrTiO$_3$, a material that is a band insulator in bulk. With increasing 3D carrier densities in these wells we observe a correlation-induced mass enhancement, followed by a transition to a correlated insulator at the highest 3D densities. We have shown that polarization-induced mobile charges in oxide quantum wells provides for a new tool to probe strong correlations associated with on-site Coulomb interactions. In particular, very large carrier densities - beyond what is possible with electric field gated structures and conventional dielectrics - can be introduced electrostatically, and without the chemical and lattice changes that accompany bulk alloying studies. Future experiments should address the precise nature (such as charge and magnetic ordering [13]) of the insulating ground state in these quantum wells, and compare the results with predictions by two-dimensional Hubbard models.

**ACKNOWLEDGEMENTS**

The authors thank Tyler Cain for help with the thin film growth, and Daniel Ouellette for help with the measurements. S.J.A., L.B., and S.S. acknowledge support by DARPA (Grant No. W911NF-12-1-0574). P.M was supported by the U.S. National Science Foundation (Grant No. DMR-1006640) and C.A.J. by the MRSEC Program of the National Science Foundation (Award No. DMR 1121053), which also supported some of the facilities that were used in this study.

**Figure Captions**

**Figure 1.** (a) Schematic showing the origin of the fixed polar charge at SrTiO$_3$/GdTiO$_3$ interfaces, compensated by ~ $3.5 \times 10^{14}$ cm$^{-2}$ mobile electrons at each interface. (b,c) HAADF-STEM images of a unit cell of SrTiO$_3$ embedded in GdTiO$_3$. The image in (b) shows a larger field of view, confirming the uniform thickness of the SrTiO$_3$ quantum well (thin dark region). The image in (c) is a magnified section of the atomic resolution image. The SrTiO$_3$ appears darker in these images because of the lower atomic number of Sr relative to Gd.

**Figure 2.** (a) Sheet resistance as a function of $T^2$ for quantum wells of SrTiO$_3$ embedded in GdTiO$_3$, with different SrTiO$_3$ quantum well thicknesses. The solid black lines are fits to Eq. (1). (b) Temperature coefficient $A$ as a function of carrier density for the samples investigated in this study and bulk La$_x$Sr$_{1-x}$TiO$_3$ from the literature (Tokura et al. [25] and Okuda et al. [24]), respectively. $A$ was obtained from fits to Eq. (1) between ~ 40 – 300 K, except for Gd$_{0.285}$Sr$_{0.715}$TiO$_3$ and Gd$_{0.56}$Sr$_{0.44}$TiO$_3$, for which it was obtained from ~ 10 – 250 K. The thickness of the SrTiO$_3$ quantum wells for the GdTiO$_3$/SrTiO$_3$/LSAT samples was between 0.4 and 8.6 nm (see supplementary material [11]) and the 3D carrier concentrations were calculated from the sheet densities as described in the text. The vertical line is a guide to the eye showing the critical carrier density for which $A$ increases as the Mott transition is approached.

**Figure 3.** Sheet resistance as a function of temperature for insulating SrTiO$_3$ quantum wells. The solid lines are the experimental data, the red dotted lines fits to an Arrhenius law, and the black dashed lines are fits to the Efros-Shklovskii law ($y = ½$).

**Figure 1**

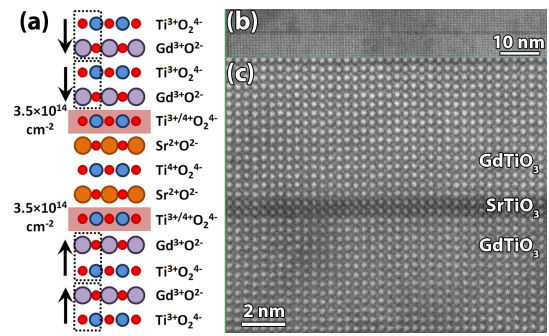

**Figure 2**

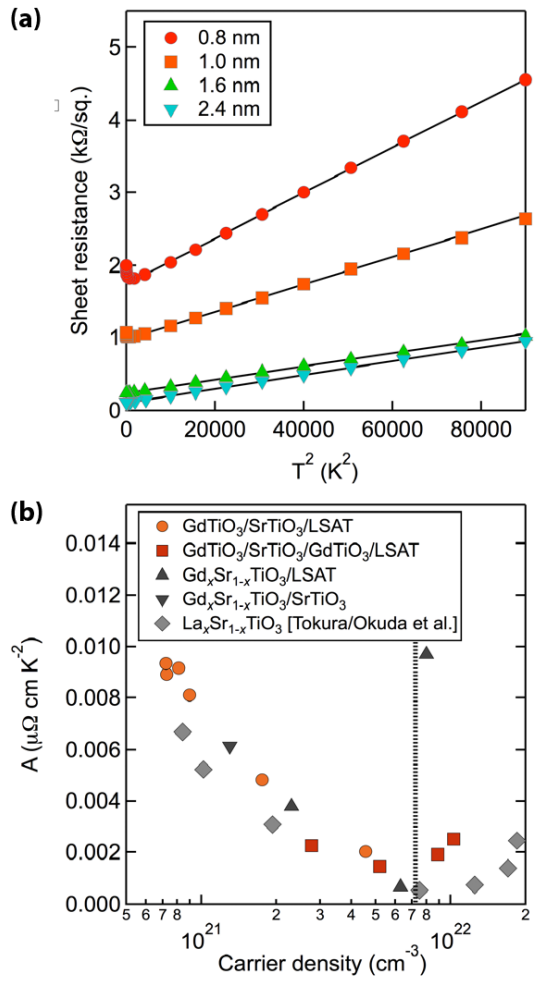

**Figure 3**

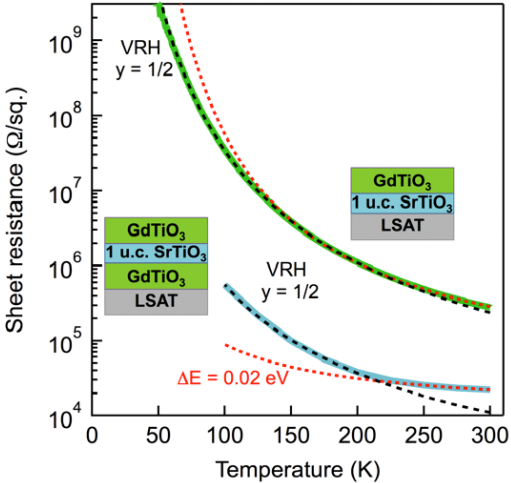